\documentclass[aps, prd, reprint, superscriptaddress, nofootinbib, bibnotes]{revtex4-2}
\usepackage{graphicx}        
\usepackage{aas_macros} 
\usepackage{amssymb,amsmath,placeins,epsfig,array,bm}
\usepackage[usenames,dvipsnames]{color}
\usepackage[breaklinks,colorlinks,urlcolor=blue,citecolor=blue,linkcolor=blue]{hyperref}
\usepackage{float}
\usepackage[T1]{fontenc}
\usepackage{times}
\usepackage{soul}

\begin{document}
\title{Peculiar Behavior of Optical Polarization in Blazar 1ES 1959+650: Role of Jet Magnetic Field and Geometry}

\author{A. Singh}
\email[]{sarchanaa019@gmail.com}
\affiliation{Homi Bhabha National Institute, Anushakti Nagar, 400094 Mumbai, India}
\affiliation{Astrophysical Sciences Division, Bhabha Atomic Research Centre, Trombay 400085 Mumbai, India}

\author{A. Tolamatti}
\affiliation{Astrophysical Sciences Division, Bhabha Atomic Research Centre, Trombay 400085 Mumbai, India}
\affiliation{Homi Bhabha National Institute, Anushakti Nagar, 400094 Mumbai, India}

\author{K. K. Singh}
\email[]{kksastro@barc.gov.in}
\affiliation{Astrophysical Sciences Division, Bhabha Atomic Research Centre, Trombay 400085 Mumbai, India}
\affiliation{Homi Bhabha National Institute, Anushakti Nagar, 400094 Mumbai, India}

\author{A. C. Gupta}
\affiliation{Aryabhatta Research Institute of Observational Sciences, Manora peak, Uttarakhand 263128, India}

\author{K. K. Yadav}
\affiliation{Astrophysical Sciences Division, Bhabha Atomic Research Centre, Trombay 400085 Mumbai, India}
\affiliation{Homi Bhabha National Institute, Anushakti Nagar, 400094 Mumbai, India}

\date{\today}

\begin{abstract}
Measurements of high degree of optical polarization from blazars provide a strong evidence 
for the optically thin synchrotron radiation from the relativistic leptons in the jet. This in 
turn characterizes the presence of a partially ordered magnetic field in the emission region. 
However, the topology of magnetic field and acceleration of electrons to ultrarelativistic energies 
within the blazar jet are poorly understood. In this work, we investigate the decade long optical light 
curve of the well-known blazar 1ES 1959+650 to probe the role of jet magnetic field in the peculiar behavior of 
the highly varying degree of linear polarization in the wavelength band 500-700 nm. The optical light curves 
in V and R-bands exhibit a strong flare without any significant change in the contemporaneous degree of 
linear polarization. However, the degree of linear polarization shows strong positive and negative/anti-correlations with the V 
and R-band fluxes measured during the epochs before and after the flare, respectively. This can be satisfactorily 
explained using the helical magnetic field and transverse shock under the framework of a relativistic jet with 
modestly varying orientations. During the flare, the fractional linear polarization is explained by the turbulent emission 
multi-zone model.
\end{abstract}


\maketitle
\section{Introduction} 
Radio-loud active galactic nuclei with their relativistic powerful plasma jets pointing at small angles to the 
Earth-based observers are referred to as the blazars \cite{Urry1995,Padovani2017,Blandford2019}. These objects 
are mainly characterized by non-thermal emission over the entire electromagnetic spectrum produced in the jet. 
The broadband radiative energy output of blazars is substantially Doppler boosted due to the geometry of their 
relativistic jets \cite{Begelman1984}. The physics behind formation, launching, collimation, and propagation 
of the blazar jets, from the central region of massive elliptical host galaxies, is not fully understood and 
remains an active area of research in astrophysics. To a large extent, it is assumed that these jets are 
launched from the vicinity of spinning supermassive black holes (SMBHs) surrounded by magnetized accretion disks of 
hot plasma \cite{Harrison2024}. The physical processes associated with the jets mainly depend on the local magnetic 
field, their plasma content and surroundings \cite{Blandford1977,Blandford1979,Blandford1982}. The magnetic field 
configuration, assumed to be closely connected to the spacetime geometry near the event horizon of SMBH, plays an important 
role in powering the relativistic jets by converting the rotational energy into Poynting energy through a confluence of 
general relativity and electrodynamics \cite{Garofalo2010}. This should lead to an imprint of the presence of SMBHs 
and their physical properties on the emission of broadband radiation from the jet plasma \cite{Sebastian2020,Tolamatti2024,Gelles2025}. 
\par
The broadband emissions from blazar jets are observed to exhibit extreme features, such as a characteristic double-hump 
structure, high degree of polarization in radio/optical/X-ray wavebands, rapid, and high amplitude flux variability 
at all wavelengths during the flaring episodes \cite{Abdo2010,Shao2019,Ajello2020,Singh2020,Shah2025}. These observational 
characteristics are explained by the acceleration and cooling of charged particles in the jet, Doppler boosting, changing 
viewing angle and apparent superluminal motion of the jet \cite{Blandford2019}. The synchrotron radiation of ultrarelativistic 
leptons (electrons/positrons) is the well-understood process for emission at radio-to-optical/X-ray part of the 
broadband spectral energy distribution (SED). This constitutes the origin of low-energy hump in the characteristic SEDs of blazars. 
The origin of high energy hump, extending from X-ray to gamma ray part of the SED, is not very clear and still remains an open problem 
in blazar research \cite{Sol2022}. The origin of flux variations at different timescales over the entire electromagnetic spectrum is not 
fully understood, indicating a  complex underlying physical process in the jet plasma. Recent studies in the literature suggest that 
macroscopic changes associated with the number density of radiating particles, or magnetic field strength in the emission region 
within the jet, or orientation of jet, can explain the occurrence of rapid flux variability during flares in blazars 
\cite{Potter2012,Joshi2016,Raiteri2017,Singh2020JHEAp,Singh2020ApSS,deMenezes2025}. Alternatively, changes in the energy 
distribution of emitting particles due to injection, acceleration, and cooling are also invoked to explain the flaring activities in 
blazars \cite{Singh2017,Bottcher2019,Shukla2020,Geng2022}.
\par
Polarimetric studies of non-thermal radiation have emerged as an important tool to understand the acceleration of radiating particles, 
magnetic field configuration, and structure in the blazar jets \cite{Zhang2019}. Signatures of polarized radiation produced due to 
synchrotron emission in radio and optical bands, measured especially during flaring episodes, provide direct probe of the properties 
of magnetic field in the blazar jet \cite{Lister2000,Abdo2010,Singh2019ApSS}. The degree of synchrotron polarization in optical band 
is measured to be higher than that at  radio frequencies \cite{Jorstad2013,Raiteri2021}. This suggests that optical emission zone  
has more ordered magnetic field than the radio emission zone within the jet. Radio emission zones are mostly related to the regions 
away from the central supermassive black hole downstream the jet \cite{Tavecchio2021}.  But, the exact configuration of jet magnetic field 
is unknown. Recent measurements of X-ray polarization from a sample of blazars suggest that shock-accelerated energetic electrons 
produce polarized synchrotron radiation \cite{Abe2025,Hu2026}. Contrary to the radio polarization, X-ray polarization can probe the acceleration region 
near the event horizon of the black hole or base of the jet. In this work, we explore the behavior of optical polarization of the blazar 
1ES 1959+650 to understand the physical processes in the jet. Recently, we have reported a long-term photo-polarimetric study of this 
source \cite{Singh2025}, and the same data have also been used to address the phenomena of photon-axion like particle coupling \cite{Huang2025}.
\par
1ES 1959+650 is a well-known BL Lac type of blazar located at redshift $\rm z = 0.047$ \cite{Schachter1993}. The low energy synchrotron hump in its broadband SED 
is observed in the UV/Optical to X-ray band \cite{Kapanadze2016}. This source has been an important target for both single and multizone emission 
processes for broadband modelling \cite{Bottcher2005,Tagliaferri2008,Patel2018,MAGIC2020,Chandra2021,Ghosal2022}. Exceptional flaring episodes have 
been detected from the blazar 1ES 1959+650 across the electromagnetic spectrum including orphan flares at TeV energies \cite{Bottcher2005,Peer2026}. 
Occurrence of orphan flares at high energies challenges the widely used single-zone leptonic models for blazar emission. Therefore, alternative 
hadronic models, including proton synchrotron radiation or synchrotron and inverse Compton processes of secondary leptons produced in photo-hadronic 
interaction, are sometimes invoked to explain the origin of orphan high energy flares \cite{Bottcher2019}. A hadronic synchrotron mirror model has been 
proposed to account for an earlier orphan TeV flare observed in the light curve of the blazar 1ES 1959+650 \cite{Bottcher2005}. In this scenario, the 
origin of the orphan TeV flare in 1ES 1959+650 can be attributed to the interaction of relativistic protons with an external photon field generated 
by leptonic synchrotron radiation. A detailed description of the past observations of this blazar can be found in \cite{Singh2025}. In the present work, 
we mainly focus on the strong positive and anti-correlations between the degree of linear polarization and optical fluxes in V and R-bands observed during 
two different epochs in the decade long temporal behavior of the blazar 1ES 1959+650. Section \ref{data} briefly describes the data set used in this work. 
Results are discussed in Section \ref{result}. Finally, we conclude in Section \ref{concl}. 
\section{Data Set}\label{data}    
In this work, we have used the publicly available archival photopolarimetric data on the blazar 1ES 1959+650 from the Spectro-Polarimeter (SPOL) at the 
Steward Observatory of the University of Arizona\footnote{\url{http://james.as.arizona.edu/~psmith/Fermi/DATA/individual.html.}}. The photopolarimetric 
observations were carried out between October 6, 2008 (MJD 54745) and June 24, 2018 (MJD 58293). The data include measurements of magnitudes in the V 
and R-bands, and the degree of linear polarization (PD) along with the angle of polarization (PA) in the wavelength range 500-700 nm \cite{Smith2009}. 
The observed magnitudes are simply converted into the energy flux densities using the standard magnitude-flux relation as described in \cite{Singh2020JHEAp}. 
\begin{figure}
      \centering
        \includegraphics[width=\columnwidth]{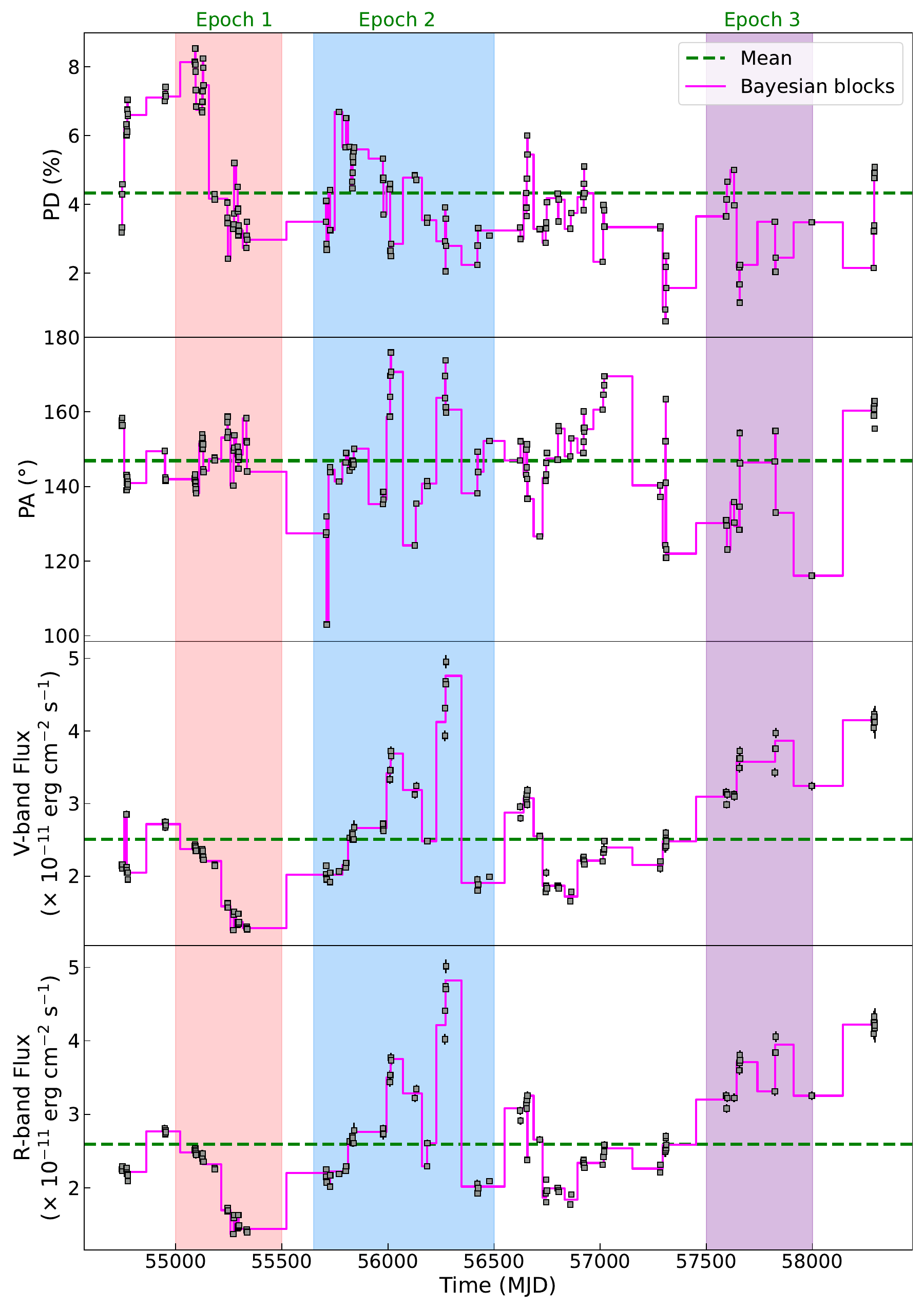}
        \caption{The photopolarimetric daily light curve of the blazar 1ES 1959+650 during the period between October 6, 2008 (MJD 54745) and 
	June 24, 2018 (MJD 58293). \textit{Epoch 1}, \textit{Epoch 2}, and \textit{Epoch 3} on the top are the time segments between two vertical lines 
	corresponding to  (MJD 55000-55500), (MJD 55650-56500), and (MJD 57500-58000) respectively. The horizontal dashed line in each panel represents the 
	mean value of corresponding physical quantity.}
    \label{fig:lc}
\end{figure}
\section{Results and Discussion}\label{result}
The long-term temporal behavior of the photometric and polarimetric emissions from the blazar 1ES 1959+650 is shown in Figure \ref{fig:lc}. 
A careful visual inspection of the photometric light curve indicates a flaring activity associated with enhanced flux densities in the V and R-bands, 
whereas no such behavior is noticed in the polarimetric light curve. In order to statistically quantify the presence of a flaring episode, we apply 
the Bayesian block algorithm\footnote{\url{https://docs.astropy.org/en/stable/api/astropy.stats.bayesian_blocks.html}} to the one-day binned light curves 
in Figure \ref{fig:lc}. A detailed mathematical description of the Bayesian block algorithm can be found in \cite{Scargle2013}. The Bayesian blocks with a 
false-alarm probability of 0.05, shown with magenta color in Figure  \ref{fig:lc}, indicate a genuine flaring activity in V and R-band light curves 
during the period MJD 55650-56500. We define this flaring episode as \emph{Epoch 2} in the decade-long light curve of the blazar 1ES 1959+650. 
Two time intervals with almost constant emission in the photometric V and R-bands during MJD 55000-55500 (pre-flare) and MJD 57500-58000 (post-flare) 
are identified as \emph{Epoch 1} and \emph{Epoch 3}, respectively.  
\subsection{Variability and Correlations}
In order to quantify the exact level of intrinsic variability, we estimate the fractional variability amplitude $(\rm F_{var})$ parameter for different 
observables in Figure \ref{fig:lc}. $\rm F_{var}$ is defined as \citep{Vaughan2003,Singh2020}:
\begin{equation}
\rm F_{var} = \sqrt{\frac{S^2 - E^2}{F^2}}
\end{equation}
and the error in $\rm F_{var}$ is given by
\begin{equation}
\rm \Delta{F_{var}} = \sqrt{\sqrt{\frac{1}{2N}}\left(\frac{E^2}{F^2 F_{var}}\right)^2 + \left(\frac{E^2}{N}\frac{1}{F}\right)^2} 
\end{equation}
where $S^2$ is the variance, $E^2$ is the mean squared error in the measurement, $F$ is the mean of the observed physical quantity, and $N$ is the 
number of data points. $F_{var}$ accounts for the uncertainty in each measurement of a physical quantity and therefore provides an estimate of the 
intrinsic variability of the source. The estimated values of  $F_{var}$ for different observables during three epochs are reported in Table \ref{tab-fvar}. 
It is observed that the degree of linear polarization is more variable during \emph{Epochs} 1 and 3 than \emph{Epoch} 2 or flaring activity in V and R-bands. 
However, the angle of polarization does not show any significant variability during all the epochs. A modest variation of approximately 
$\rm \pm10^{\circ}$ is observed in PA during Epoch 3. This is negligibly small as compared to the large PA swings measured during major flaring 
episodes in blazars \cite{Abdo2010,Blinov2016}. It may be noted that despite X-ray and bright TeV-flares observed from the source during Epoch 3 
variation in PA is not significant \cite{Chandra2021,MAGIC2020}.  
\par
\begin{figure}
	\centering
	\includegraphics[width=1.0\columnwidth]{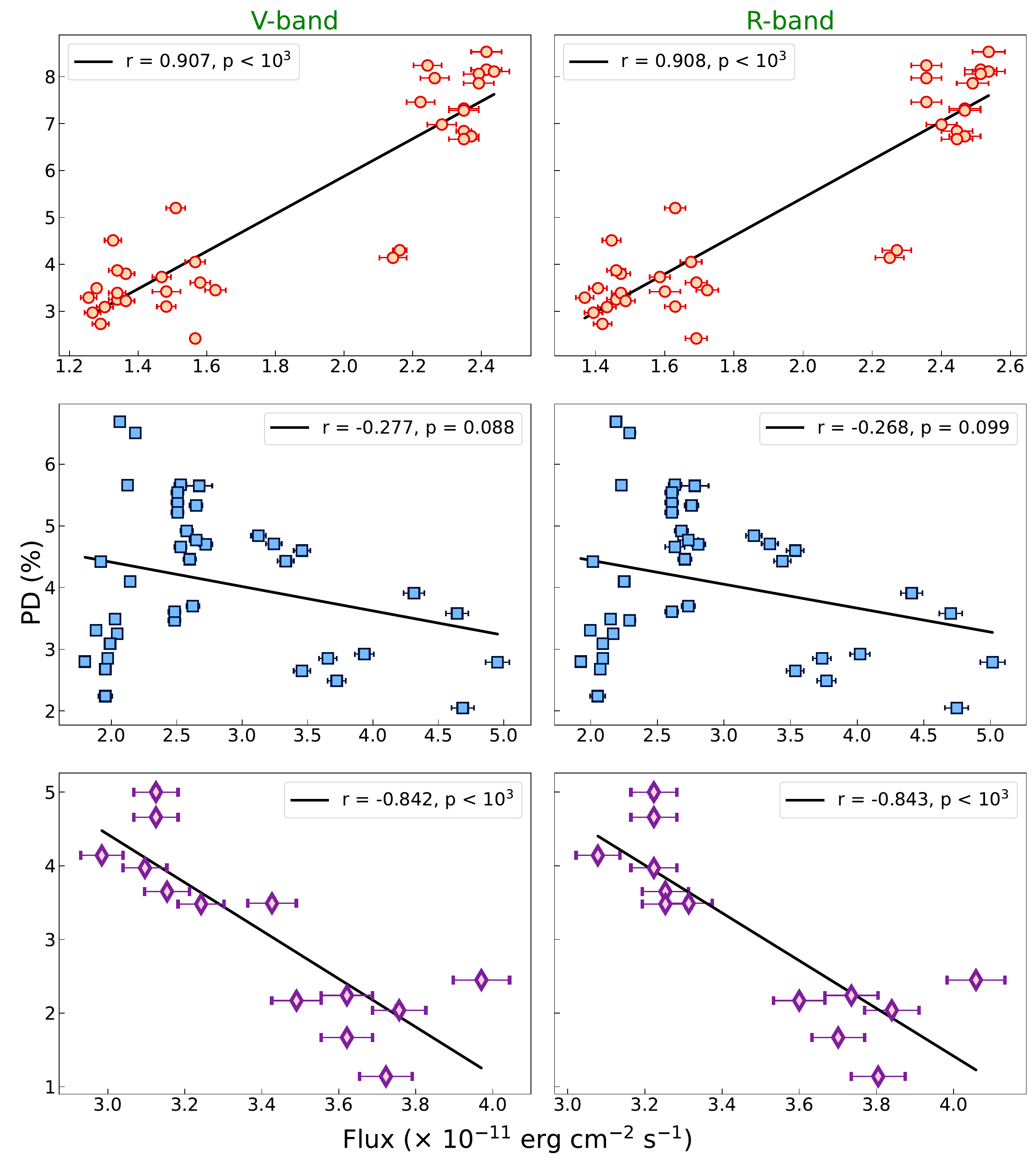}
	\caption{Scatter plots for the degree of linear polarization and optical fluxes in V and R bands during Epoch 1 (top), Epoch 2 (middle), and 
	Epoch 3 (bottom) for the blazar 1ES~1959+650. Lines represent a linear fit to the data.}
\label{fig:corr}
\end{figure}

\begin{figure}
	\centering
	\includegraphics[width=1.0\columnwidth]{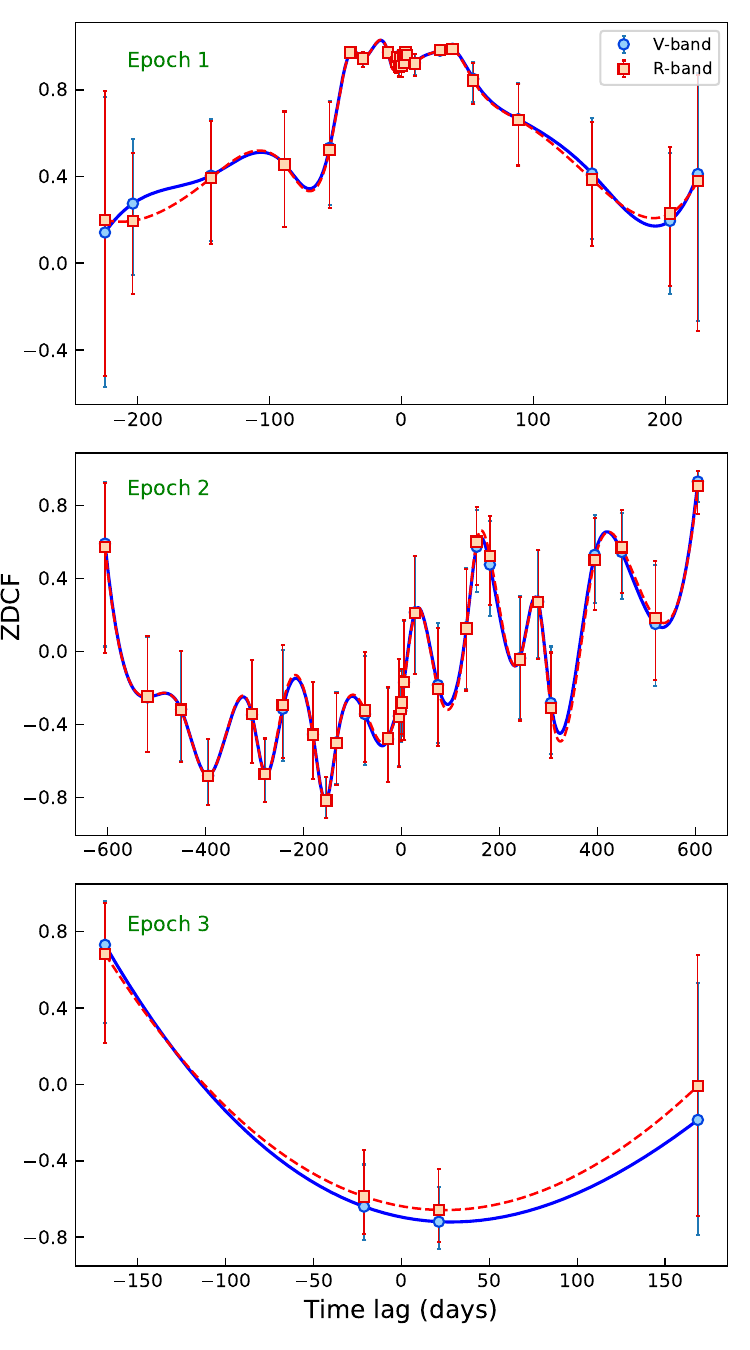}
	\caption{Analysis curves of discrete correlation function vs time-lag between the degree of linear polarization (P) and 
		 optical fluxes (V and R-bands) during Epoch 1 (top), Epoch 2 (middle), and Epoch 3 (bottom).}
\label{fig:zdcf}
\end{figure}

 
To investigate the possible correlation between the polarization degree and contemporaneous fluxes in the V and R-bands, we calculate the Pearson's correlation 
coefficient (r) and Spearman's rank correlation coefficient ($\rho$) during the three epochs. The scatter plots along with $r$ and corresponding $p-values$ are 
shown in Figure \ref{fig:corr}. The null hypothesis that there is no correlation between the two observables is rejected if $p < 0.05$.
\setlength{\tabcolsep}{5pt}
\begin{table}
\caption{Summary of estimated values of $\rm F_{var} \pm \Delta F_{var}$ in percentage for various measurements during Epoch 1, Epoch 2, and Epoch 3.}
\label{tab-fvar}
\begin{tabular}{lccc}
\hline\hline
Observable 	&Epoch 1	&Epoch 2	& Epoch 3\\	
\hline\hline
V-Flux 		&25.51 $\pm$ 1.81  &30.02 $\pm$ 2.00  &8.58 $\pm$1.85 \\\hline
R-Flux 		&23.63 $\pm$ 1.91  &28.66 $\pm$ 2.04  &8.47 $\pm$ 1.85 \\\hline
Pol. Deg.	&39.74 $\pm$ 1.08  &29.12 $\pm$ 1.17  &37.60 $\pm$ 1.53\\\hline
Pol. Ang.	&3.68  $\pm$ 0.28  &10.02 $\pm$ 0.27  &8.27 $\pm$ 0.40 \\\hline
\end{tabular}
\end{table}
From the scatter plots at top (Epoch 1) and bottom (Epoch 3), it is evident that Epoch 1 (top) and Epoch 3 (bottom) respectively indicate strong positive 
and negative/anti-correlations between the degree of polarization and optical fluxes. However, apparently no/weak anti-correlation is observed 
between the polarimetric and photometric measurements in the middle plots (Epoch 2). These findings pertaining to the linear correlations between the 
polarization degree and optical fluxes are also confirmed by the estimated values of $\rho$ (given in Table \ref{tab-corr}) corresponding to the three 
epochs. In order to further explore the plausible correlations between the polarization and optical measurements, we have also estimate the z-transformed 
discrete correlation function (ZDCF) using a publicly available code described in \cite{Alexander1997}. The results of the ZDCF analysis for all the three 
epochs are reported in Figure \ref{fig:zdcf}. For Epoch 1, the DCF peaks around zero time-lag indicating a significant positive correlation between the polarization 
and optical fluxes. Multiple small peaks and dips in the DCF during the flare period (Epoch 2) imply no significant correlation between polarization and optical 
measurements. For Epoch 3, a dip around zero time lag in the DCF indicates a clear anti-correlation between the degree of linear polarization and optical fluxes 
in V and R-band.
\par
\begin{figure}
	\centering
	\includegraphics[width=1.0\columnwidth]{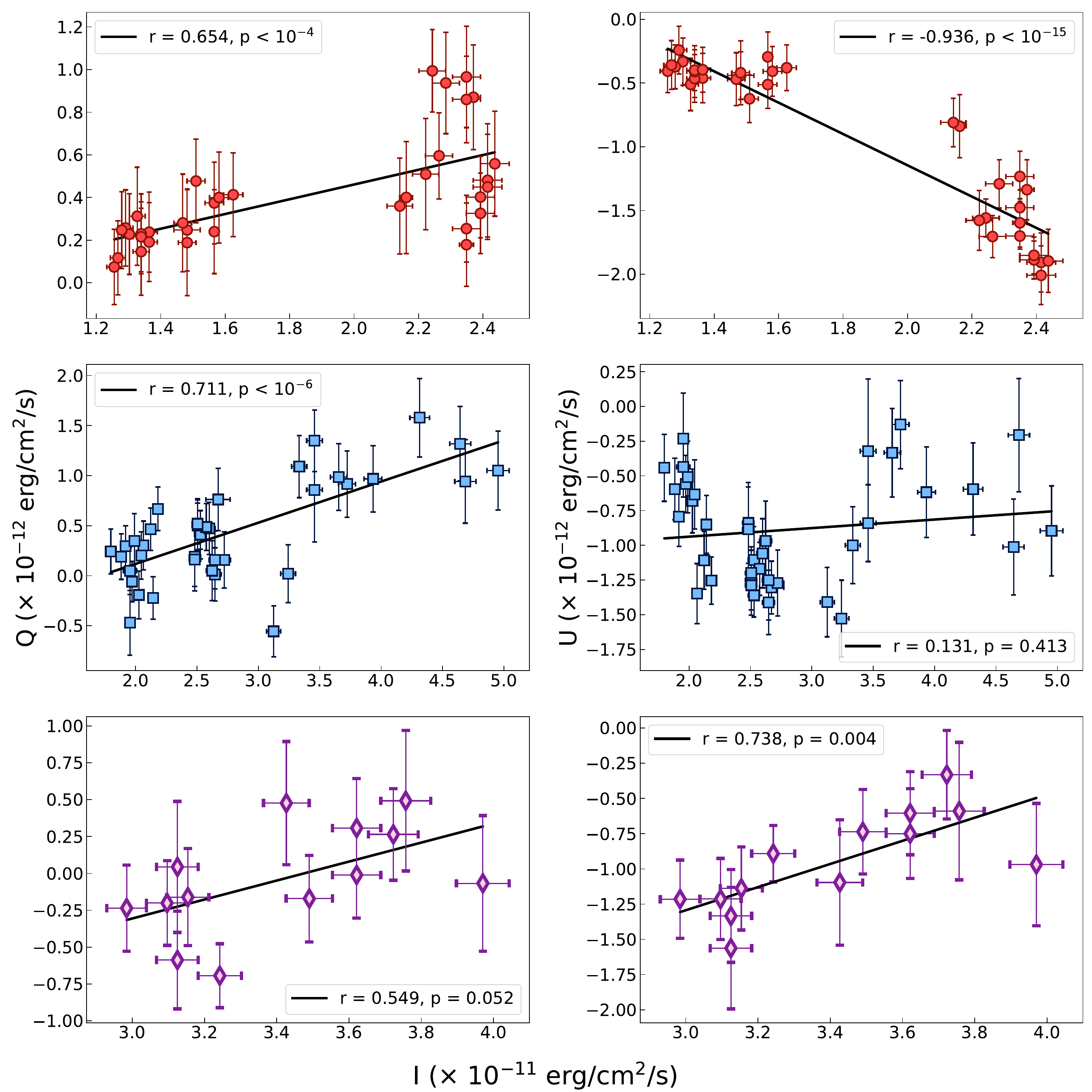}
	\caption{Scatter plots for the absolute Stokes parameters Q and U vs. I during Epoch 1 (top), Epoch 2 (middle), and Epoch 3 (bottom) for the 
			blazar 1ES~1959+650 with solid lines representing a linear dependence.}
\label{fig:stokes}
\end{figure}

The scatter plots for absolute Stokes parameters (Q, U, I) obtained from the linear polarization measurements are also shown in Figure \ref{fig:stokes}.  
Assuming that I is approximately given by the flux density in V-band,  r-values are estimated corresponding to Q vs. I and U vs. I during the three epochs. 
Strong positive linear-correlations are obtained between Q and I during Epochs 1 and 2, whereas Epoch 3 exhibits a moderate positive-correlation. 
However, a strong anti-correlation and positive-correlation are observed between U and I for Epoch 1 and Epoch 3 respectively. And, Epoch 2 shows a very 
weak positive-correlation between U and I. The corresponding $\rho$-values are reported in Table \ref{tab-stokes}.      

\setlength{\tabcolsep}{5pt}
\begin{table}
	\caption{Summary of Spearman's correlation coefficient $\rho$ (p-value) between the degree of polarization and optical fluxes.}
\label{tab-corr}
\begin{tabular}{lccc}
\hline\hline
	 	&Epoch 1	&Epoch 2	& Epoch 3\\	
\hline\hline
V-band 		&0.831 ($< 10^{-3}$) &-0.045 (0.785) &-0.837($< 10^{-3}$)  \\\hline
R-band 		&0.824 ($< 10^{-3}$) &-0.047 (0.773) &-0.841($< 10^{-3}$)  \\\hline
\end{tabular}
\end{table}
\setlength{\tabcolsep}{5pt}
\begin{table}
	\caption{Summary of Spearman's correlation coefficient $\rho$ (p-value) for the absolute Stoke's parameters.}
\label{tab-stokes}
\begin{tabular}{lcc}
\hline\hline
	 	&Q vs. I	&U vs. I\\	
\hline\hline
Epoch 1		&0.699 ($< 10^{-6}$) &-0.870 ($< 10^{-11}$)  \\\hline
Epoch 2		&0.606 ($< 10^{-5}$) &-0.108 ($< 10^{-3}$)  \\\hline
Epoch 3		&0.606 ($< 10^{-3}$) &-0.813 ($< 10^{-3}$)  \\\hline
\end{tabular}
\end{table}

\subsection{Possible Physical Scenario}
The degree of linear optical polarization in the blazar 1ES 1959+650 is observed to show a negligible correlation with the contemporaneous 
emissions in the V and R-bands measured during the flaring state (Epoch 2) of the source. A strong positive correlation during Epoch 1, before 
the optical flare, indicates strengthened ordering or orientation of magnetic field lines in the emission region. The post-flare anti-correlation 
between the degree of polarization and fluxes in V and R-bands implies a misalignment of the emission region with large scale magnetic field of the 
blazar jet. As the behavior of the degree of polarization with the fluxes in the V and R-bands is almost similar, we consider only V-band measurements in 
the following scenarios for the optical emission region in the jet of blazar 1ES 1959+650. 
\begin{figure}
\centering
\includegraphics[width=1.0\columnwidth]{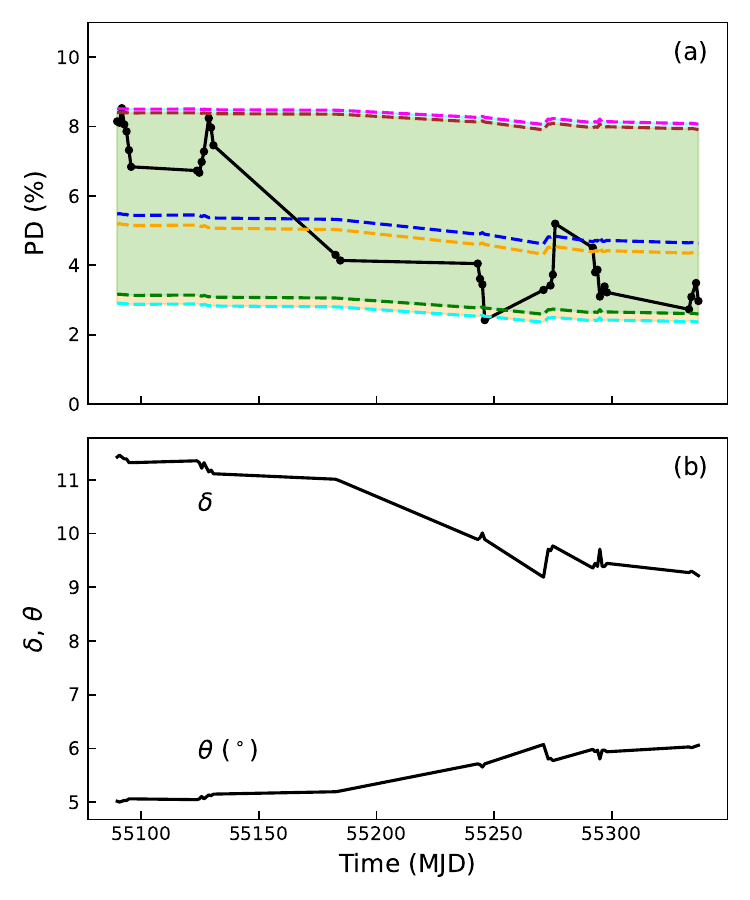}
	\caption{Results from the helical magnetic field and transverse shock scenarios during Epoch 1. 
	{\bf (a)} The solid black line represents the observed degree of polarization. The dashed magenta, blue and 
	green lines represent the time-dependent behavior of polarization predicted by the helical magnetic field model 
	for $\rm \theta = 5^{\circ}$, $9.5^{\circ}$, \& $14^{\circ}$ respectively and $\rm \Gamma_b = 12$. 
	The dashed brown, orange and cyan lines show the polarization behavior predicted by the transverse shock model 
	for $\rm \theta = 5^{\circ}$, $9.5^{\circ}$, \& $14^{\circ}$ respectively and $\Gamma_b = 12$. The green 
	shaded region represents the full range of model uncertainty (Equation \ref{eqn:mod}). {\bf (b)} 
	Time dependence of $\rm \delta$ and $\rm \theta$ for $\rm \Gamma_b = 12$, and a minimum viewing angle of $\rm 5^{\circ}$.}
\label{fig:hm-epoch1}
\end{figure}
\begin{figure}
\centering
\includegraphics[width=1.0\columnwidth]{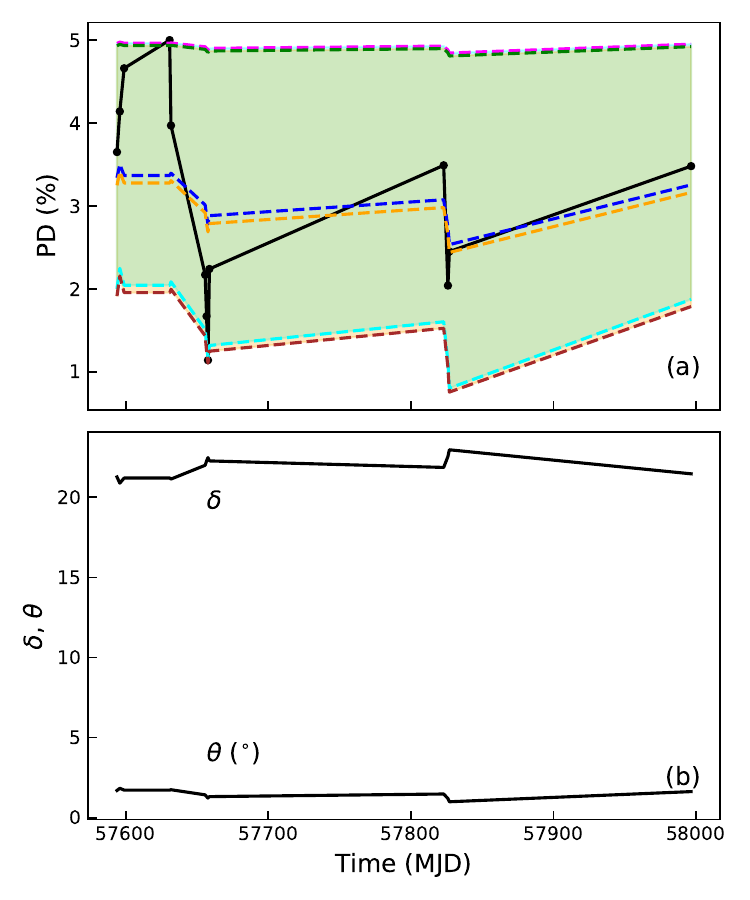}
	\caption{Results from the helical magnetic field and transverse shock scenarios during Epoch 3. 
	{\bf (a)} The solid black line represents the observed degree of polarization. The dashed magenta, blue and 
	cyan lines represent the time-dependent behavior of polarization predicted by the helical magnetic field model 
	for $\rm \theta = 4^{\circ}, 2^{\circ}, \& 1^{\circ}$ respectively and $\rm \Gamma_b = 12$. 
	The dashed green, orange and brown lines show the polarization behavior predicted by the transverse shock model 
	for $\rm \theta = 4^{\circ}, 2^{\circ}, \& 1^{\circ}$ respectively and $\Gamma_b = 12$. The green 
	shaded region indicates the full range of model uncertainty (Equation \ref{eqn:mod}). {\bf (b)} 
	Time dependence of $\rm \delta$ and $\rm \theta$ for $\rm \Gamma_b = 12$, and a minimum viewing angle of $\rm 1^{\circ}$.}
\label{fig:hm-epoch3}
\end{figure}
\begin{itemize}
	\item{\bf Helical Magnetic Field:} The blazar emission zone is an unresolved region in the parsec-scale jet with its highly magnetized base 
		at the launching site near the event horizon of the SMBH. The non-thermal emission from this highly collimated 
		magnetized plasma jet is relativistically Doppler boosted due to its orientation along the line of sight of the observer. The Doppler 
		factor is given by \cite{Urry1995}
		\begin{equation}\label{eqn:delta}
			\rm \delta = \frac{1}{\Gamma_b (1-\beta cos\theta)}
		\end{equation}	
where $\rm \Gamma_b = (1-\beta^2)^{-1/2}$ is the bulk Lorentz factor of the blob emitting non-thermal radiation, $\rm \beta$ is the velocity of the 
		emission region in units of the speed of light, and $\rm \theta$ is the jet viewing angle. For a continuous blazar jet, the Doppler 
		boosting factor is $\rm \delta^{2 + \alpha}$, where $\rm \alpha$ is the mean spectral index of synchrotron radiation emitted from 
		the blob. From Equation \ref{eqn:delta}, it is obvious that the observed flux density ($\rm F$) can vary if $\delta$ changes due 
		to varying $\rm \theta$ even if the intrinsic emission from the source remains constant. If $\rm F_{max}$ is the highest measured flux 
		density from the source and $\rm \delta_{max}$ being the corresponding Doppler factor, we have 
		\begin{equation}\label{eqn:delta-time}
			\rm \delta = \delta_{max} \left(\frac{F}{F_{max}}\right)^{1/{(2 + \alpha)}}
		\end{equation}
		Therefore, the time-dependent behavior of $\rm \delta$ can be derived using $\rm F$ from the light curve in 
		Equation \ref{eqn:delta-time} for a given $\rm \Gamma_b$ and minimum viewing angle. This can be used to infer 
		the time-dependent trend of $\rm \theta$ also. 
\par
High resolution radio observations suggest that blazar jets are neither straight nor steady structures \cite{Kellermann2004,Marscher2011,Bloom2013}. 
Changes observed in the Lorentz factor, viewing angle, and azimuthal angle suggest bending of the jet with varying speed. Assuming that rotation of 
central SMBH along with its magnetized accretion disk generates a helical magnetic field structure that is carried outward along the 
jet \cite{Blandford1977}. In this scenario, the synchrotron emission from the plasma blob is modulated by the time-dependent viewing angle in the 
jet rest frame due to the helical magnetic field embedded within the jet leading to a measurable variations in the degree of polarization. 
Therefore, the degree of linear polarization for optically thin synchrotron emission from a relativistic jet carrying large-scale helical magnetic
field can be  expressed as \cite{Lyutikov2005,Raiteri2013}
\begin{equation}\label{eqn:mod}
 \rm P_{hel} = P_{max} sin^2  \theta^\prime 
\end{equation}
where $\rm P_{max}$ is the maximum degree of polarization and $\rm \theta^\prime$ is the instantaneous aberrated viewing angle in the jet rest frame. 
$\rm \theta^\prime$ is related to the viewing angle $\rm \theta$ in observer's frame through the Lorentz transformation as   
\begin{equation}
 \rm     sin \theta^\prime = \frac{sin \theta}{\Gamma_b (1-\beta cos \theta)} 
\end{equation}
The time-dependent behavior of the degree of linear polarization ($\rm P_{hel}$), predicted by the scenario of jet having large scale helical magnetic field, 
		is derived by varying $\rm \theta$ while fixing $\rm \Gamma_b = 12$ from the literature \cite{Das2025}. The model predictions are found to 
		be broadly consistent with the measurements during Epoch 1 and Epoch 3 with $\rm P_{max} \sim 8.5\%$ and 5\%, respectively (Figure \ref{fig:lc}). 
		Results from the comparison for Epoch 1 and Epoch 3 are depicted in Figure \ref{fig:hm-epoch1}(a) and \ref{fig:hm-epoch3}(a) respectively.
		It is evident that the observed variations in the degree of polarization restrict the viewing angle between $\rm 5^{\circ}$ and $ 14^{\circ}$ during 
		Epoch 1. However, the viewing angle lies in the range $\rm 1^{\circ} - 4^{\circ}$ during Epoch 3. 
		The time-evolution of $\rm \delta$ and $\rm \theta$ as inferred from the observations are shown in Figure \ref{fig:hm-epoch1}(b) and 
		\ref{fig:hm-epoch3}(b) for Epoch 1 and  Epoch 3, respectively. These trends correspond to $\rm \Gamma_b = 12$ and minimum viewing angle 
		of $\rm 5^{\circ}$ (Epoch 1) and $\rm 1^{\circ}$ (Epoch 3). For the helical jet magnetic field, large apparent changes in the 
		jet direction can be interpreted as small bends which are boosted by projection \cite{Murphy2013,Gabuzda2018}. At small viewing angles, 
		the magnetic field projection gives rise to more symmetric polarization profiles, which reduces the net degree of polarization. In contrast, 
		a larger viewing angle, the helical magnetic field structure is viewed more side-on producing asymmetry in the projected polarization profiles 
		and therefore resulting in a higher degree of polarization. Thus, the helical jet magnetic field scenario broadly explains the average behavior of 
		the degree of linear polarization and its anti-correlation with the fluxes in V and R-bands measured from the blazar 1ES 1959+650.  

	\item{\bf Transverse Shock:} We also investigate the role of shocks, an important intrinsic mechanism in blazar jets, to understand the complex nature of 
		polarization. In the so called shock-in-jet model, the chaotic or random magnetic field is compressed by the passage of a shock wave \cite{Hughes1985}. 
		A transverse shock moving downstream in the jet compresses the relativistic jet plasma. This compression orders the magnetic field lines within the shocked 
		emission region due to flux freezing and thus leads to the variations in the degree of linear polarization. In this scenario, the observed degree of 
		polarization can be expressed as \cite{Hagen2008}
		\begin{equation}
     		\rm P_{sw} \approx \frac{\alpha + 1}{\alpha + \frac{5}{3}} \frac{(1 - \eta^{-2}) \sin^{2}\theta^\prime}{2 - (1 - \eta^{-2}) \sin^{2}\theta^\prime}
\end{equation}
where $\rm \alpha$ is the synchrotron spectral index and $\rm \eta$ is the shock compression ratio. $\rm \eta$ represents the ratio of density in the shocked region to 
the unshocked region. For transverse shock waves, the compression axis coincides with the jet axis. Therefore, the  viewing angle of the shock is same as the jet viewing 
angle in the rest frame ($\rm \theta^\prime$). We use the derived values of $\rm \alpha$ from the optical photomteric measurements in V and R-bands 
(as defined in \cite{Singh2025}) to predict the degree of polarization by the transverse shock wave model ($\rm P_{sw}$) for $\rm \Gamma_b = 12$. 
Comparisons of the predicted degree of polarization under transverse shock scenario with the measured values for Epoch 1 and Epoch 3 are shown in 
Figure \ref{fig:hm-epoch1}(a) and \ref{fig:hm-epoch3}(a), respectively. The corresponding best agreement values of $\rm \eta$ for found to be  $\sim 1.128$ (Epoch 1) 
and $\sim 1.070$ (Epoch 3). It is also evident from the figures that the predictions from the transverse shock model are also compatible with that from the helical 
magnetic field model. In both the scenarios, decreasing minimum viewing angle leads to an increase in the optical flux and the degree of polarization reduced variability. 
This explains the positive correlation between the optical flux and the degree of polarization observed during Epoch 1. Conversely, increasing minimum viewing angle results 
in a decrease in optical flux and a higher polarization degree. This can broadly account for the anti-correlation during Epoch 3. 
\end{itemize}
\begin{figure}
\centering
\includegraphics[width=1.0\columnwidth]{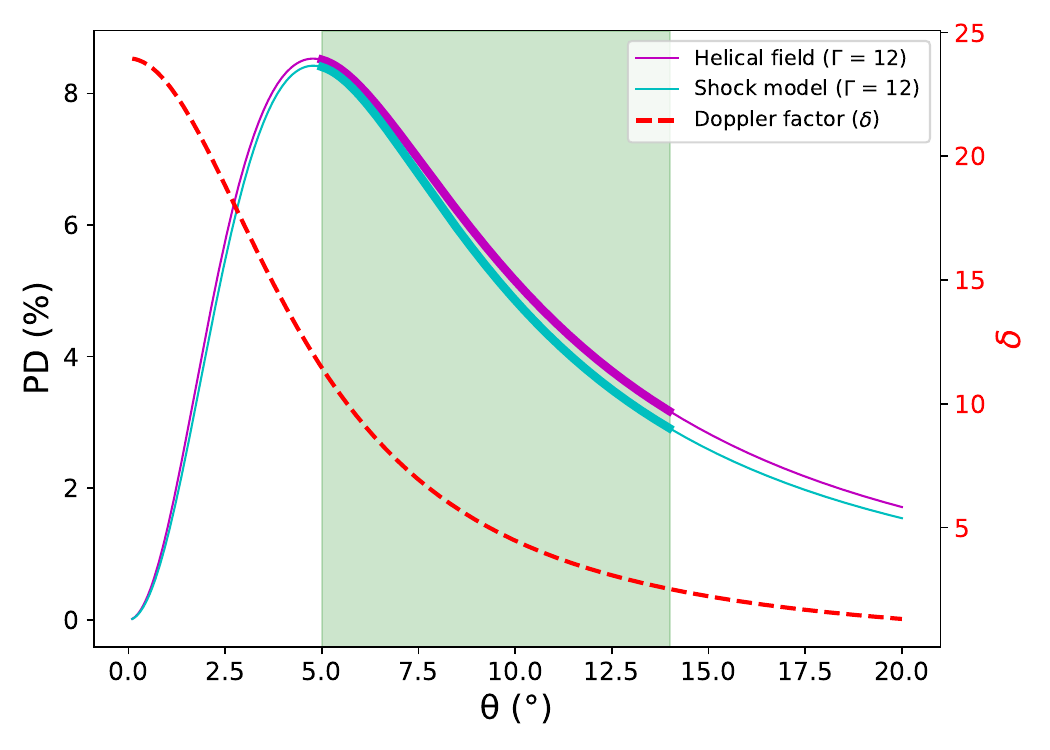}
\caption{Epoch 1:Degree of linear polarization predicted by the helical magnetic field and transverse shock models as a function of the jet viewing angle ($\rm \theta$) 
          for $\rm \Gamma_b = 12$. The dashed red line represents the variation of Doppler factor as a function of $\theta$ for fixed $\Gamma_b = 12$ and minimum 
	  viewing angle of $\rm 5^{\circ}$. The vertical green shaded region represents the range of model calculations.}
	\label{fig:epoch1-model}
\end{figure}
\begin{figure}
\centering
\includegraphics[width=1.0\columnwidth]{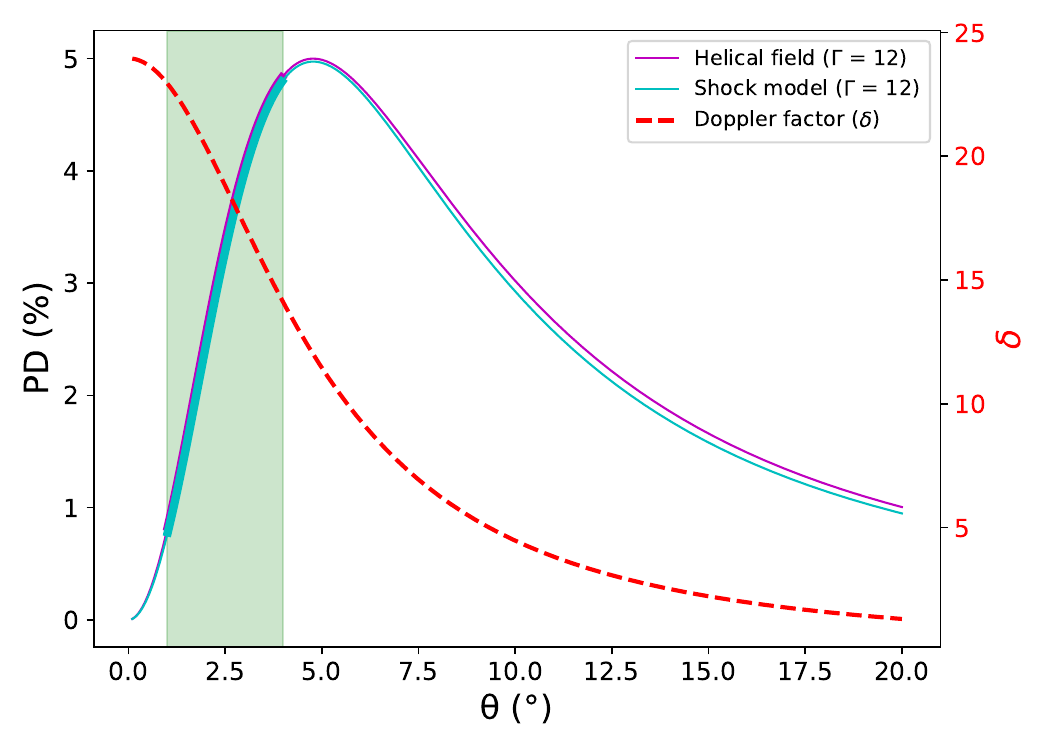}
	\caption{Epoch 3: Same as Figure \ref{fig:epoch1-model} with a minimum viewing angle of $\rm 1^{\circ}$.}
\label{fig:epoch3-model}
\end{figure}
\par
Illustrations of the variation of the degree of linear polarization, predicted by the helical magnetic field and the transverse shock models, as a function of the 
jet viewing angle ($\rm \theta$) during Epoch 1 and Epoch 3 are presented in Figures \ref{fig:epoch1-model} and \ref{fig:epoch3-model}, respectively. 
In both the cases, we fix the bulk Lorentz factor $\rm \Gamma_b= 12$ considering the minimum viewing angles of $\rm 5^{\circ}$ (Epoch 1) and $\rm 1^{\circ}$ (Epoch 3). 
It is observed that the predicted degree of polarization increases rapidly with increasing viewing angle, attains a peak at $\rm \theta = 5^{\circ}$, and 
subsequently starts decreasing gradually. In the present case, the emission region in the jet is observed within a limited range of viewing angle, i.e., 
$\rm 5^\circ \leq \theta \leq 14^\circ$ during Epoch 1 and $\rm 1^\circ \leq \theta \leq 4^\circ$ during Epoch 3. For Epoch 1, the observationally relevant 
$\rm \theta$ range lies entirely within the descending segment of the polarization curve, whereas for Epoch 3, it lies entirely within the ascending part of 
the polarization curve. This indicates that an increase in the viewing angle leads to a reduction in the Doppler factor $\delta$, and consequently, to a decrease in 
the observed flux. However, over the same interval, the degree of linear polarization will decrease for the descending segment (Epoch 1) and increase for the 
ascending segment (Epoch 3). These are consistent with the expected behavior of a helical magnetic field configuration and transverse shock scenarios 
accounting for the observed correlation (Epoch 1) and anti-correlations (Epoch 2) between the degree of linear polarization and flux densities in V and R-bands.
\begin{figure}
\centering
\includegraphics[width=1.0\columnwidth]{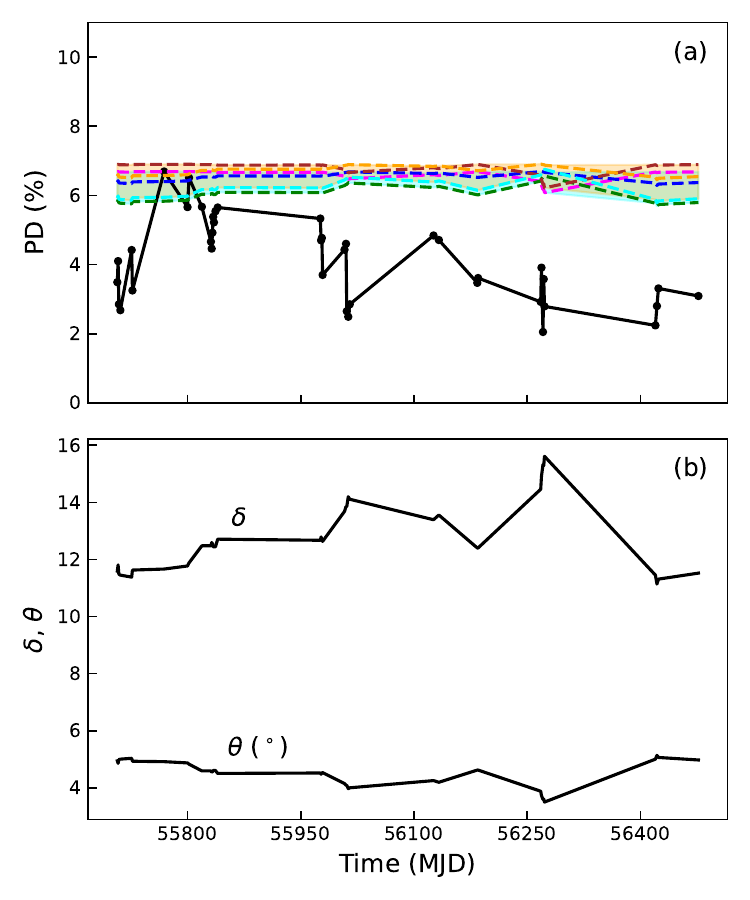}
\caption{Predictions from the helical jet magnetic field and transverse shock models (for a modest variation of $\theta$) and measured degree of 
	polarization during Epoch 2. {\bf (a)} The solid black line indicates the observed degree of polarization. The dashed magenta, blue 
	and cyan lines represent the polarization behavior predicted by the helical magnetic field model for $\rm \theta = 5.5^{\circ}, 4.5^{\circ},
	\& 3.5^{\circ}$ respectively and $\Gamma_b = 12$. The dashed green, orange and pink lines show the polarization behavior predicted by 
	the transverse shock model $\rm \theta = 5.5^{\circ}, 4.5^{\circ}, \& 3.5^{\circ}$ respectively and $\rm \Gamma_b = 12$. 
	{\bf (b)} Long-term time dependent behavior of $\rm \delta$ and $\rm \theta$ for $\rm \Gamma_b = 12$ and a minimum viewing angle of  
	$\rm 3.5^{\circ}$.}
\label{fig:hm-epoch2}
\end{figure}
\par
However both the scenarios, helical jet magnetic field and transverse shock, are unable to account for the behavior of the degree of polarization during the 
flaring activity associated with the optical emission of the source (Epoch 2). A comparison of the model prediction trends and observed degree of polarization 
during Epoch 2 is presented in Figure \ref{fig:hm-epoch2}. As discussed in Section A, both these observables exhibit a very weak anti-correlation or no correlation 
during the flare period in optical emission. This may be explained within the framework of the Turbulent Emission Multi-Zone (TEMZ) model \cite{Marscher2014, 
Pandey2022,Singh2025}. The optical emission is produced from numerous turbulent cells in the jet. And, the optical flare occurs when one or more turbulent cells 
with enhanced electron density or a stronger magnetic field pass through a standing shock. The shock compresses the plasma and accelerates the radiating 
charged particles (electrons) to higher energies. This leads to a temporary increase in the synchrotron emissivity in the optical band due to modest local 
enhancements in particle or magnetic energy density. However, the overall magnetic field structure remains largely disordered. As the net polarization is a 
vector sum of emission from many cells, the addition of emission from multiple zones with different magnetic field orientations can dilute or randomly modify 
the resultant polarization. As a result, large optical flux variations can temporarily occur without a systematic increase or decrease in the degree 
of polarization. The long-term average behavior of the degree polarization from the blazar 1ES 1959+650 is consistent with the scenario of stochastic turbulence 
dominating the magnetic field configuration rather than a globally ordered field \cite{Singh2025}. 
\par
\citet{Raiteri2013} have interpreted the long-term flux and polarization behaviour of the blazar BL Lacertae in the framework of helical magnetic field 
and transverse shock models. Authors have concluded that the geometrical interpretation of the flux variability in combination with the above models can 
succesfully explain the long-term behavior of polarization from  BL Lacertae. For the same blazar, \citet{Gaur2014} explained the strong anti-correlation 
observed between V-band flux and the degree of polarization by involving the helical jet magnetic field and single transverse shock. They concluded that the 
temporary anti-correlation between the degree of polarization and V-band flux can be attributed to the shock-in-spiral-jet picture for a relatively low Lorentz 
factor of the shock. Recent polarimetric observations of the blazar 1ES 1959+650 with an average degree of X-ray polarization above 10\% also supports the 
presence of helical jet magnetic field or turbulent plasma governed by the stochastic processes in the jet \cite{Pacciani2025}. The X-ray polarization measured 
from the blazar 1ES 1959+650 favors shock acceleration of electrons in the jet \cite{Bharathan2024}. A low degree of X-ray polarization from this source is also 
attributed to turbulence in the relativistic jet flow \cite{Errando2024}.
\section{Conclusions}\label{concl} 
In the present work, we have studied the behavior of optical polarization from the blazar 1ES 1959+650 using a decade long observations with the SPOL 
at the Steward Observatory. The contemporaneous photometric emissions in V and R-bands exhibit a high emission state during MJD 55650-56500. Based on 
this flaring episode, the optical light curve of the source is divided into three epochs namely pre-flare (Epoch 1), flare (Epoch 2) and post-flare (Epoch 3). 
The degree of linear polarization is observed to show a high variability during pre- and post-flare epochs and a relatively low variability during the flare period. 
Further, the polarization degree shows strong positive and negative-correlations with the V and R-band fluxes both during the pre- and post-flare periods, respectively. 
However, a very weak anti-correlation is observed between the polarization degree and V and R-band fluxes during the flaring episode. This contradicting and erratic 
behavior of optical polarization can be explained by considering an appropriate geometry of the jet and two plausible physical scenarios for the emission. 
The jet of blazar 1ES 1959+650 contains different emission regions with different viewing angles. During the pre-and post-flare periods, the viewing angle of the 
optical emission regions varies in the range $\rm 5^{\circ}$ -- $14^{\circ}$ and $\rm 1^{\circ}$ -- $4^{\circ}$, respectively and the bulk Lorentz factor of the jet 
is fixed at 12. The observed variations in the long-term optical emission of the blazar 1ES 1959+650 can be purely attributed to the time-dependent orientations 
of the emission region. Modest variations in the viewing angle are sufficient to reproduce the long-term behavior of the observed flux in V and R-bands. 
And, the correponding variations in the degree of optical linear polarization can be explained by two alternate physical scenarios pertaining to the helical jet magnetic field and the transverse shocks. Both the models not only yield the polarization values compared to the observed ones but also qualitatively 
predict the major trends observed in the polarization data during the pre- and post flare epochs. Nevertheless, both the scenarios fail to account for the observed 
polarization during the optical flare. The observed behavior during the flare period can be attributed to the turbulent emission from multi-zones in the jet. 
Therefore, a more complex physical model involving turbulence, multi-zone emission, and relativistic particle acceleration, is needed to explain the full 
long-term behavior of polarization from the blazar-like sources.
\section*{Data Availability}
The data that support the findings of this article are publicly available \cite{Smith2009}.
\\\\
\textbf{Acknowledgments} 
Authors thank the anonymous referee for his/her insightful comments and constructive suggestions. Data from the Steward Observatory spectropolarimetric 
monitoring project were used. This program is supported by Fermi Guest Investigator grants NNX08AW56G, NNX09AU10G, NNX12AO93G, and NNX15AU81G.

\bibliography{paper}

\end{document}